\documentclass[%
preprint,
showkeys,
nofootinbib,
nobibnotes,
 amsmath,amssymb,
 aps,
pra,
 longbibliography,
 lengthcheck,%
]{revtex4-1}

\usepackage{graphicx}
\usepackage{dcolumn}
\usepackage{bm}
\usepackage{hyperref}
\usepackage{multirow} 
\usepackage{relsize}

\newcommand{\kF}{k_{\rm F}}

\newcommand{\be}{\begin{equation}}
\newcommand{\ee}{\end{equation}}
\def\bea{\begin{eqnarray}}
\def\eea{\end{eqnarray}}

\usepackage{amsthm}

\begin{document}


\title{Reply to "Comment on 'Kinetic theory for a mobile impurity in a degenerate Tonks-Girardeau gas'"}

\author{O. Gamayun$^{1,2,3}$}
\author{O. Lychkovskiy$^{4}$}
\author{V. Cheianov$^{1,3}$}
\affiliation{
$^1$
Instituut-Lorentz, Universiteit Leiden, P.O. Box 9506, 2300 RA Leiden, Netherlands,
}
\affiliation{$^2$Bogolyubov Institute for Theoretical Physics, 14-b Metrolohichna str., Kyiv 03680, Ukraine,}
\affiliation{%
$^3$Lancaster University, Physics Department, Lancaster LA1 4YB, UK,
}%
\affiliation{$^4$Russian Quantum Center, Novaya St. 100A, Skolkovo, Moscow Region, 143025, Russia.}
%


\begin{abstract}
In our recent paper [Phys. Rev. E 90, 032132 (2014)] we have studied the dynamics of a mobile impurity particle weakly interacting with the Tonks-Girardeau gas and pulled by a small external force, $F$. Working in the regime when the thermodynamic limit is taken prior to the small force limit, we have found that the Bloch oscillations of the impurity velocity are absent in the case of a light impurity. Further, we have argued that for a light impurity the steady state drift velocity, $V_D$, remains finite in the limit $F\rightarrow 0$. These results are in contradiction with earlier works by Gangardt, Kamenev and Schecter [Phys. Rev. Lett. 102, 070402 (2009), Annals of Physics 327, 639–670 (2012)]. One of us (OL) has conjectured [Phys. Rev. A 91, 040101 (2015)] that the central assumption of these works -- the adiabaticity of the dynamics -- can  break down in the thermodynamic limit. In the preceding Comment [Phys. Rev. E 92, 016101 (2015)]  Schecter, Gangardt and Kamenev have argued against this conjecture and in support of the existence of Bloch oscillations and linearity of $V_D(F)$. They have suggested that the ground state of the impurity-fluid system is a quasi-bound state and that this is sufficient to ensure adiabaticity in the thermodynamic limit. Their analytical argument is based on a certain truncation of the Hilbert space of the system.  We argue that extending the results and intuition based on their truncated model on the original many-body problem lacks justification.

\end{abstract}

\maketitle


\date{\today}

The preceding Comment  \cite{schecter2014comment} discusses the discrepancy between the results of Refs. \cite{Gangardt2009,schecter2012dynamics,schecter2012critical} on the one side and the results of Ref. \cite{Gamayun2014} on the other side concerning the dynamics of a mobile impurity particle pulled through a one-dimensional (1D) fluid by a small constant force. The results in question are as follows: it has been predicted in  \cite{Gangardt2009,schecter2012dynamics,schecter2012critical} that the impurity experiences Bloch oscillations (in the absence of any external periodic potential) superimposed on the drift and that the drift velocity, $V_D$, is linear in  force, $F$. This behavior has been predicted to be fairly general, the only validity condition being the smoothness of the lower edge of the impurity-fluid spectrum \cite{schecter2012critical}. On the other hand, it has been argued in Ref. \cite{Gamayun2014} that the oscillations do not occur for the impurity in the Tonks-Girardeau gas and $V_D(F\rightarrow 0)$ remains finite, provided the impurity is lighter than a gas particle.


First we would like to stress that there is no fundamental contradiction between the results obtained with the Boltzmann equation
and a hypothetic possibility to observe Bloch oscillations. The reason for this is as follows. The Boltzmann equation is robust and applicable to a generic initial state, whether pure or thermal, while the  Bloch oscillations have been predicted in Refs. \cite{Gangardt2009,schecter2012dynamics} for a very special choice of the initial state, namely, the ground state. Considering the absence of the gap in the spectrum, the preparation of the system in the ground state might be quite difficult experimentally. However, with this caveat in mind, question remains if the Bloch oscillations could be observed even if such preparation is done. In this reply we present our arguments as to why we think this has not been convincingly demonstrated.

The Bloch oscillations have been derived in \cite{Gangardt2009,schecter2012dynamics,schecter2012critical} by a straightforward application of the adiabatic theorem to the impurity-fluid system. In one dimension the spectral edge of the latter is always periodic in momentum which readily leads to oscillations. Undoubtedly, this reasoning holds for a finite system provided the driving force is sufficiently small. However a key ambiguity is the exact meaning of "sufficiently small" in view of the fact that the many-body system under consideration is gapless \cite{Lychkovskiy2014}. To be more specific, the question is how the critical force at which the adiabaticity breaks down scales in the thermodynamic limit. The reasoning of \cite{Gangardt2009,schecter2012dynamics,schecter2012critical} is justified only if the critical force does not vanish in this limit.

The aforementioned crucial question, which has not been addressed in Refs. \cite{Gangardt2009,schecter2012dynamics,schecter2012critical} as well as in Ref.  \cite{Gamayun2014}, is the focus of the Comment. The authors of the Comment argue that the critical force is finite in the thermodynamic limit.

This principle claim of the Comment contradicts a common intuition about gapless many-body systems, see e.g. a discussion in section 5.2.3 of the textbook \cite{balian2007microphysics}. Further still, the critical force has been found to vanish in several specific gapless many-body systems \cite{polkovnikov2008breakdown,altland2008many,altland2009nonadiabaticity}, while we are not aware of any specific system in which the opposite would be explicitly demonstrated.

The authors of the Comment consider a toy model with a truncated Hilbert space (in the fermionic language, only one particle-hole excitation is allowed in this model) in order to justify their claim. In contrast to the original many-body system, the ground state of this truncated model is separated from the continuum of excited states by a finite gap. As a consequence, the critical force in the truncated model is finite. 

The finite gap in the truncated model can be associated with a bound state. The authors of the Comment admit that the gap is absent in the original many-body system. Nevertheless, they point to the power-law correlation function in the original system and argue that  a ``quasi-bound'' ground state with binding energy equal to the gap persists and ensures adiabaticity as well as linearity of $V_D(F)$. We do not see, however, a direct logical relation between the power-law behavior of correlation function on the one hand and adiabaticity of quantum evolution and linearity of the response on the other.

The authors of the Comment state that "the one-hole bound-state solution is in {\it quantitative} agreement  with the available exact results". As a substantiation of  this statement, they compare the bound state energy in the truncated model and in the integrable many-body model (fluid in the Tonks-Girardeau limit, mass of the impurity $m$ equal to the mass of the fluid particle) in the limit $\gamma \ll 1$, $\gamma$ being a dimensionless impurity-fluid coupling, and in the vicinity of the total momentum $P=\kF$. Indeed, at this specific total momentum the truncated model gives the correct ground state energy $\frac{\kF^2}{2 m}$. We emphasize, however, that this concordance is destroyed for momenta away from $P=\kF$. E.g. for $P=0$
\be
E_g =\frac{\kF^2}{\pi^2 m}\gamma-\frac{\kF^2}{8\pi^2 m}\gamma^2+ O(\gamma^3),
\ee
see \cite{mcguire1965interacting}, while
\be
E_g^{\rm tr} = \frac{\kF^2}{\pi^2 m}\gamma-\Delta + O(\Delta^2)~~{\rm with}~~ \Delta=\frac{2\kF^2}{m} e^{-2\pi^2/\gamma}.
\ee
The discrepancy $|E_g-E_g^{tr}|$ is on the order of $O(\gamma^2)$ and thus is much larger than the exponentially small binding energy $\Delta$. This observation is also valid in the nonintegrable case where the ground state energy can be calculated perturbatively \cite{lamacraft2009dispersion}. Thus, the truncated model introduces a nonperturbative exponentially small correction to the ground state energy of the system but misses much larger perturbative contributions. For this reason it can {\it not} be regarded as being in ``quantitative agreement  with the available exact results''.

The above estimate elucidates a physical reason to doubt the existence of the  quasi-bound state: The fluctuation of the kinetic energy of the impurity-fluid system is $O(\gamma^2)$ and hence is likely to completely destroy the alleged quasi-bound state with an exponentially small binding energy.


Finally, we comment on the drift velocity of the impurity as a function of the driving force. The authors of the Comment  focus their attention on the heavy impurity case when it comes to the drift velocity, but do not mention the light impurity case. In fact, in the former case our results can be reconciled, as discussed both in Ref. \cite{Gamayun2014} and in the Comment. On the other hand, in the latter case the discrepancy is dramatic:  According to Refs.  \cite{schecter2012dynamics,schecter2012critical} the drift velocity always vanishes at $ F\rightarrow 0$, while according to Ref. \cite{Gamayun2014} it remains finite in this limit. The root cause of this discrepancy is closely related to the presumed reason for the absence of Bloch oscillations in the light impurity case \cite{Gamayun2014}.

To conclude, we believe that the conclusions and intuition obtained on the basis of the truncated model suggested in the Comment  cannot be justifiably extended to the original impurity-fluid system. Thus an important question  whether the adiabatic driving in the considered many-body system exists in the thermodynamic limit remains  open. Answering this question is necessary for understanding the driven dynamics of a mobile impurity in a 1D quantum fluid and, in particular, in resolving the controversy between Refs. \cite{Gangardt2009,schecter2012dynamics,schecter2012critical} and Ref. \cite{Gamayun2014}.  We plan to address this problem in our future research.

{\em Acknowledgements.} The work was supported by the ERC grant 279738-NEDFOQ.

\bibliography{D:/Work/QM/Bibs/1D,D:/Work/QM/Bibs/LZ_and_adiabaticity}

\providecommand{\noopsort}[1]{}\providecommand{\singleletter}[1]{#1}
\begin{thebibliography}{12}%
\makeatletter
\providecommand \@ifxundefined [1]{%
 \@ifx{#1\undefined}
}%
\providecommand \@ifnum [1]{%
 \ifnum #1\expandafter \@firstoftwo
 \else \expandafter \@secondoftwo
 \fi
}%
\providecommand \@ifx [1]{%
 \ifx #1\expandafter \@firstoftwo
 \else \expandafter \@secondoftwo
 \fi
}%
\providecommand \natexlab [1]{#1}%
\providecommand \enquote  [1]{``#1''}%
\providecommand \bibnamefont  [1]{#1}%
\providecommand \bibfnamefont [1]{#1}%
\providecommand \citenamefont [1]{#1}%
\providecommand \href@noop [0]{\@secondoftwo}%
\providecommand \href [0]{\begingroup \@sanitize@url \@href}%
\providecommand \@href[1]{\@@startlink{#1}\@@href}%
\providecommand \@@href[1]{\endgroup#1\@@endlink}%
\providecommand \@sanitize@url [0]{\catcode `\\12\catcode `\$12\catcode
  `\&12\catcode `\#12\catcode `\^12\catcode `\_12\catcode `\%12\relax}%
\providecommand \@@startlink[1]{}%
\providecommand \@@endlink[0]{}%
\providecommand \url  [0]{\begingroup\@sanitize@url \@url }%
\providecommand \@url [1]{\endgroup\@href {#1}{\urlprefix }}%
\providecommand \urlprefix  [0]{URL }%
\providecommand \Eprint [0]{\href }%
\providecommand \doibase [0]{http://dx.doi.org/}%
\providecommand \selectlanguage [0]{\@gobble}%
\providecommand \bibinfo  [0]{\@secondoftwo}%
\providecommand \bibfield  [0]{\@secondoftwo}%
\providecommand \translation [1]{[#1]}%
\providecommand \BibitemOpen [0]{}%
\providecommand \bibitemStop [0]{}%
\providecommand \bibitemNoStop [0]{.\EOS\space}%
\providecommand \EOS [0]{\spacefactor3000\relax}%
\providecommand \BibitemShut  [1]{\csname bibitem#1\endcsname}%
\let\auto@bib@innerbib\@empty
\bibitem [{\citenamefont {Schecter}\ \emph {et~al.}(2015)\citenamefont
  {Schecter}, \citenamefont {Gangardt},\ and\ \citenamefont
  {Kamenev}}]{schecter2014comment}%
  \BibitemOpen
  \bibfield  {author} {\bibinfo {author} {\bibfnamefont {Michael}\ \bibnamefont
  {Schecter}}, \bibinfo {author} {\bibfnamefont {Dimitri~M.}\ \bibnamefont
  {Gangardt}}, \ and\ \bibinfo {author} {\bibfnamefont {Alex}\ \bibnamefont
  {Kamenev}},\ }\bibfield  {title} {\enquote {\bibinfo {title} {Comment on
  ``{Kinetic} theory for a mobile impurity in a degenerate tonks-girardeau
  gas''},}\ }\href {\doibase 10.1103/PhysRevE.92.016101} {\bibfield  {journal}
  {\bibinfo  {journal} {Phys. Rev. E}\ }\textbf {\bibinfo {volume} {92}},\
  \bibinfo {pages} {016101} (\bibinfo {year} {2015})}\BibitemShut {NoStop}%
\bibitem [{\citenamefont {Gangardt}\ and\ \citenamefont
  {Kamenev}(2009)}]{Gangardt2009}%
  \BibitemOpen
  \bibfield  {author} {\bibinfo {author} {\bibfnamefont {D.~M.}\ \bibnamefont
  {Gangardt}}\ and\ \bibinfo {author} {\bibfnamefont {A.}~\bibnamefont
  {Kamenev}},\ }\bibfield  {title} {\enquote {\bibinfo {title} {{Bloch}
  oscillations in a one-dimensional spinor gas},}\ }\href {\doibase
  10.1103/PhysRevLett.102.070402} {\bibfield  {journal} {\bibinfo  {journal}
  {Phys. Rev. Lett.}\ }\textbf {\bibinfo {volume} {102}},\ \bibinfo {pages}
  {070402} (\bibinfo {year} {2009})}\BibitemShut {NoStop}%
\bibitem [{\citenamefont {Schecter}\ \emph
  {et~al.}(2012{\natexlab{a}})\citenamefont {Schecter}, \citenamefont
  {Gangardt},\ and\ \citenamefont {Kamenev}}]{schecter2012dynamics}%
  \BibitemOpen
  \bibfield  {author} {\bibinfo {author} {\bibfnamefont {M.}~\bibnamefont
  {Schecter}}, \bibinfo {author} {\bibfnamefont {D.~M.}\ \bibnamefont
  {Gangardt}}, \ and\ \bibinfo {author} {\bibfnamefont {A.}~\bibnamefont
  {Kamenev}},\ }\bibfield  {title} {\enquote {\bibinfo {title} {Dynamics and
  {Bloch} oscillations of mobile impurities in one-dimensional quantum
  liquids},}\ }\href@noop {} {\bibfield  {journal} {\bibinfo  {journal} {Annals
  of Physics}\ }\textbf {\bibinfo {volume} {327}},\ \bibinfo {pages} {639--670}
  (\bibinfo {year} {2012}{\natexlab{a}})}\BibitemShut {NoStop}%
\bibitem [{\citenamefont {Schecter}\ \emph
  {et~al.}(2012{\natexlab{b}})\citenamefont {Schecter}, \citenamefont
  {Kamenev}, \citenamefont {Gangardt},\ and\ \citenamefont
  {Lamacraft}}]{schecter2012critical}%
  \BibitemOpen
  \bibfield  {author} {\bibinfo {author} {\bibfnamefont {M.}~\bibnamefont
  {Schecter}}, \bibinfo {author} {\bibfnamefont {A.}~\bibnamefont {Kamenev}},
  \bibinfo {author} {\bibfnamefont {D.~M.}\ \bibnamefont {Gangardt}}, \ and\
  \bibinfo {author} {\bibfnamefont {A.}~\bibnamefont {Lamacraft}},\ }\bibfield
  {title} {\enquote {\bibinfo {title} {Critical velocity of a mobile impurity
  in one-dimensional quantum liquids},}\ }\href {\doibase
  10.1103/PhysRevLett.108.207001} {\bibfield  {journal} {\bibinfo  {journal}
  {Phys. Rev. Lett.}\ }\textbf {\bibinfo {volume} {108}},\ \bibinfo {pages}
  {207001} (\bibinfo {year} {2012}{\natexlab{b}})}\BibitemShut {NoStop}%
\bibitem [{\citenamefont {Gamayun}\ \emph {et~al.}(2014)\citenamefont
  {Gamayun}, \citenamefont {Lychkovskiy},\ and\ \citenamefont
  {Cheianov}}]{Gamayun2014}%
  \BibitemOpen
  \bibfield  {author} {\bibinfo {author} {\bibfnamefont {O.}~\bibnamefont
  {Gamayun}}, \bibinfo {author} {\bibfnamefont {O.}~\bibnamefont
  {Lychkovskiy}}, \ and\ \bibinfo {author} {\bibfnamefont {V.}~\bibnamefont
  {Cheianov}},\ }\bibfield  {title} {\enquote {\bibinfo {title} {Kinetic theory
  for a mobile impurity in a degenerate {Tonks-Girardeau} gas},}\ }\href@noop
  {} {\bibfield  {journal} {\bibinfo  {journal} {Phys. Rev. E}\ }\textbf
  {\bibinfo {volume} {90}},\ \bibinfo {pages} {032132} (\bibinfo {year}
  {2014})}\BibitemShut {NoStop}%
\bibitem [{\citenamefont {Lychkovskiy}(2015)}]{Lychkovskiy2014}%
  \BibitemOpen
  \bibfield  {author} {\bibinfo {author} {\bibfnamefont {O.}~\bibnamefont
  {Lychkovskiy}},\ }\bibfield  {title} {\enquote {\bibinfo {title} {Perpetual
  motion and driven dynamics of a mobile impurity in a quantum fluid},}\ }\href
  {\doibase 10.1103/PhysRevA.91.040101} {\bibfield  {journal} {\bibinfo
  {journal} {Phys. Rev. A}\ }\textbf {\bibinfo {volume} {91}},\ \bibinfo
  {pages} {040101} (\bibinfo {year} {2015})}\BibitemShut {NoStop}%
\bibitem [{\citenamefont {Balian}\ \emph {et~al.}(2007)\citenamefont {Balian},
  \citenamefont {Gregg},\ and\ \citenamefont {ter
  Haar}}]{balian2007microphysics}%
  \BibitemOpen
  \bibfield  {author} {\bibinfo {author} {\bibfnamefont {R.}~\bibnamefont
  {Balian}}, \bibinfo {author} {\bibfnamefont {J.~F.}\ \bibnamefont {Gregg}}, \
  and\ \bibinfo {author} {\bibfnamefont {D.}~\bibnamefont {ter Haar}},\
  }\href@noop {} {\emph {\bibinfo {title} {From microphysics to
  macrophysics}}}\ (\bibinfo  {publisher} {Springer},\ \bibinfo {year}
  {2007})\BibitemShut {NoStop}%
\bibitem [{\citenamefont {Polkovnikov}\ and\ \citenamefont
  {Gritsev}(2008)}]{polkovnikov2008breakdown}%
  \BibitemOpen
  \bibfield  {author} {\bibinfo {author} {\bibfnamefont {A.}~\bibnamefont
  {Polkovnikov}}\ and\ \bibinfo {author} {\bibfnamefont {V.}~\bibnamefont
  {Gritsev}},\ }\bibfield  {title} {\enquote {\bibinfo {title} {Breakdown of
  the adiabatic limit in low-dimensional gapless systems},}\ }\href@noop {}
  {\bibfield  {journal} {\bibinfo  {journal} {Nature Phys.}\ }\textbf {\bibinfo
  {volume} {4}},\ \bibinfo {pages} {477--481} (\bibinfo {year}
  {2008})}\BibitemShut {NoStop}%
\bibitem [{\citenamefont {Altland}\ and\ \citenamefont
  {Gurarie}(2008)}]{altland2008many}%
  \BibitemOpen
  \bibfield  {author} {\bibinfo {author} {\bibfnamefont {A.}~\bibnamefont
  {Altland}}\ and\ \bibinfo {author} {\bibfnamefont {V.}~\bibnamefont
  {Gurarie}},\ }\bibfield  {title} {\enquote {\bibinfo {title} {Many body
  generalization of the {Landau-Zener} problem},}\ }\href@noop {} {\bibfield
  {journal} {\bibinfo  {journal} {Phys. Rev. Lett.}\ }\textbf {\bibinfo
  {volume} {100}},\ \bibinfo {pages} {063602} (\bibinfo {year}
  {2008})}\BibitemShut {NoStop}%
\bibitem [{\citenamefont {Altland}\ \emph {et~al.}(2009)\citenamefont
  {Altland}, \citenamefont {Gurarie}, \citenamefont {Kriecherbauer},\ and\
  \citenamefont {Polkovnikov}}]{altland2009nonadiabaticity}%
  \BibitemOpen
  \bibfield  {author} {\bibinfo {author} {\bibfnamefont {A.}~\bibnamefont
  {Altland}}, \bibinfo {author} {\bibfnamefont {V.}~\bibnamefont {Gurarie}},
  \bibinfo {author} {\bibfnamefont {T.}~\bibnamefont {Kriecherbauer}}, \ and\
  \bibinfo {author} {\bibfnamefont {A.}~\bibnamefont {Polkovnikov}},\
  }\bibfield  {title} {\enquote {\bibinfo {title} {Nonadiabaticity and large
  fluctuations in a many-particle {Landau-Zener} problem},}\ }\href@noop {}
  {\bibfield  {journal} {\bibinfo  {journal} {Phys. Rev. A}\ }\textbf {\bibinfo
  {volume} {79}},\ \bibinfo {pages} {042703} (\bibinfo {year}
  {2009})}\BibitemShut {NoStop}%
\bibitem [{\citenamefont {McGuire}(1965)}]{mcguire1965interacting}%
  \BibitemOpen
  \bibfield  {author} {\bibinfo {author} {\bibfnamefont {J.~B.}\ \bibnamefont
  {McGuire}},\ }\bibfield  {title} {\enquote {\bibinfo {title} {Interacting
  fermions in one dimension. {I}. {Repulsive} potential},}\ }\href@noop {}
  {\bibfield  {journal} {\bibinfo  {journal} {J. Math. Phys.}\ }\textbf
  {\bibinfo {volume} {6}},\ \bibinfo {pages} {432} (\bibinfo {year}
  {1965})}\BibitemShut {NoStop}%
\bibitem [{\citenamefont {Lamacraft}(2009)}]{lamacraft2009dispersion}%
  \BibitemOpen
  \bibfield  {author} {\bibinfo {author} {\bibfnamefont {A.}~\bibnamefont
  {Lamacraft}},\ }\bibfield  {title} {\enquote {\bibinfo {title} {Dispersion
  relation and spectral function of an impurity in a one-dimensional quantum
  liquid},}\ }\href@noop {} {\bibfield  {journal} {\bibinfo  {journal} {Phys.
  Rev. B}\ }\textbf {\bibinfo {volume} {79}},\ \bibinfo {pages} {241105}
  (\bibinfo {year} {2009})}\BibitemShut {NoStop}%
\end{thebibliography}%

\end{document}